\documentclass[baaa]{baaa}

\usepackage[pdftex]{hyperref}
\usepackage{subfigure}
\usepackage{natbib}
\usepackage{helvet,soul}
\usepackage[font=small]{caption}
\usepackage{float}

\usepackage{soul}

\contriblanguage{1}

\contribtype{1}

\thematicarea{5}

\received{\ldots}
\accepted{\ldots}

\title{Understanding the chemistry of the young stellar object G29.862$-$0.0044}

\titlerunning{The chemistry in YSO G29}

\author{
N.C. Martinez\inst{1,2},
S. Paron\inst{1},
D. Mast\inst{3},
M.E. Ortega\inst{1},
A. Petriella\inst{1}
\&
C. Fariña\inst{4,5}
}

\authorrunning{Martinez N. C. et al.}

\contact{nmartinez@iafe.uba.ar}

\institute{
Instituto de Astronom{\'\i}a y F{\'\i}sica del Espacio, CONICET--UBA, Argentina
\and
Universidad de Buenos Aires, Facultad de Ciencias Exactas y Naturales, Departamento de F\'isica, Buenos Aires, Argentina
\and
Observatorio Astron\'omico de C\'ordoba, UNC, Argentina
\and
Isaac Newton Group of Telescopes, E38700, La Palma, España
\and
Instituto de Astrof\'{\i}sica de Canarias (IAC) and Universidad de La Laguna, Dpto. Astrof\'{\i}sica, España
}

\resumen{En trabajos previos, hemos investigado la región de formación estelar G29.96$-$0.02 donde se encuentra el objeto estelar joven masivo (MYSO, por sus siglas en inglés) G29.862$-$0.0044 (de ahora en adelante G29) embebido en un núcleo molecular caliente. En uno de ellos, de carácter multiespectral, que incluyó datos obtenidos con el Atacama Submillimeter Telescope Experiment (ASTE), Atacama Large Millimeter Array (ALMA) y NIRI-Gemini, se investigó a G29 en distintas escalas espaciales. Sin embargo, su intrigante morfología revelada en el infrarrojo cercano junto con la distribución del gas molecular de su entorno, hacen que el escenario en el cual dicha fuente evoluciona esté muy lejos de ser comprendido. Este trabajo incorpora el análisis de la emisión de varias líneas moleculares adquiridas con ALMA, que no habían sido previamente examinadas (ej.:~CH$_{3}$OH, HC$_{3}$N, H$_{ 2}$CO, C$^{34}$S, H$_{2}$CS) y se presenta una nueva determinación de la temperatura. Adicionalmente, se presenta un resultado preliminar obtenido de observaciones espectroscópicas realizadas con el instrumento NIFS en Gemini y en continuo de radio obtenidas con el Karl G. Jansky Very Large Array (JVLA). Esta investigación permitrá llevar a cabo un minucioso estudio químico de la región, que contribuirá a avanzar en la comprensión de los procesos físicos involucrados en la formación estelar de alta masa.}

\abstract{In previous works, we have investigated the star-forming region G29.96$-$0.02 where the massive young stellar object (MYSO) G29.862$-$0.0044 (hereafter G29) is embedded in a hot molecular core.
In one of them, of multiwavelength nature, using data from the Atacama Submillimeter Telescope Experiment (ASTE), data from the Atacama Large Millimeter Array (ALMA), and photometric data from NIRI-Gemini, G29 was investigated at different spatial scales. However, the intriguing morphology of G29 in the near-infrared, together with the distribution of the associated molecular gas, reveals that the star-formation scenario is far from being understood. This work incorporates the analysis of the emission of several molecular lines acquired with ALMA that were not previously examined (eg.,~CH$_{3}$OH, HC$_{3}$N, H$_{ 2}$CO, C$^{34}$S, H$_{2}$CS) as well as a new determination of the temperature of the region. Additionally, we present the progress of results obtained through new observations in the near-infrared, in this case spectroscopic, using NIFS-Gemini, and in radio continnum obtained with the Karl G. Jansky Very Large Array (JVLA). This research allows us to carry out a detailed chemical study of the region, which will contribute to the understanding of the physical processes involved in the high-mass star formation.}


\begin{document}

\maketitle
\section{Introduction}\label{S_intro}

Star-forming processes have a deep influence on the chemistry of the molecular cores in which the young stellar objects (YSOs) are embedded and in their surroundings (e.g. \citealt{jorgen20}).

\begin{figure}
    \centering
    \includegraphics[width=7.2cm]{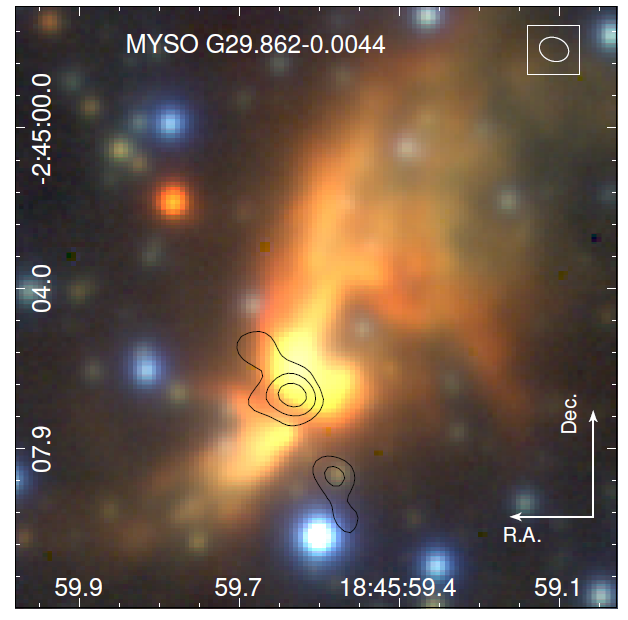}
    \caption{Three-color image of the region toward G29 with the JHK broad-band emission presented in blue, green, and red, respectively, obtained with Gemini-NIRI \citep{areal20}. Black contours represent the continuum at 1.3 mm obtained with ALMA at levels 5, 10, and 20 mJy~beam$^{-1}$ (rms = 1.5 mJy~beam$^{-1}$). The ALMA beam size is 0$.\!\!^{\prime\prime}$78$\times$0$.\!\!^{\prime\prime}$60~and is indicated in the upper right corner.}
    \label{g29}
\end{figure}

In \citet{areal20}, YSO\,G29.862$-$0.0044 (G29) was investigated at core and clump spatial scales ($<$0.2 and $\sim$0.5 pc, respectively), and even though it was performed a deep study about several physical processes, some issues remain open. Among them, it is still uncertain whether G29 consists of a single YSO or multiple sources contributing to the observed structures in near-infrared (NIR) emission (see Fig.\,\ref{g29}). Another important topic that deserves more study is the chemistry, which can give us
important information about the processes that are ongoing in the region. 

Figure\,\ref{g29} presents the region where G29 is located. The image, composed of three colors, shows the JHKs emission in blue, green, and red respectively, obtained with Gemini-NIRI (see \citealt{areal20}). This NIR emission is quite asymmetrical: it extends broad and open toward the north with two possible jet-like structures separated by diffuse emission while at the south, another jet feature remains small and sharp. Both nebulosities are apart by a dark lane, probably a disk of material masking a central protostar.
Lying proximate to the dark lane and located almost at the center of the NIR emission, a core of cold dust mapped at 1.3 mm with the Atacama Large Millimeter Array (ALMA) is discerned (see black contours in Fig.\,\ref{g29}).  
Toward this region, ciano radical (CN) and methyl cyanide (CH$_{3}$CN) emissions were previously examined \citep{areal20,areal21}. 

In this preliminary study, we present new chemical species observed in the region where the YSO is located, previously unidentified. Additionally, based on new multiwavelength observations, we introduce a new physical analysis aimed at further describing the star-forming scenario related to G29.

\section{Observations and data}

To perform the analysis of G29, we used a high-quality set of data and observations at different wavelengths.

Radio continuum observations at 3\,cm were carried out using the Karl G. Jansky Very Large Array (JVLA) in A configuration (Project 22A-063; PI: M. Ortega). The field of view of such observations uncovers the whole G29 structure with a beam of 0$.\!\!^{\prime\prime}$42$\times$0$.\!\!^{\prime\prime}$17. Additionally, we used data from ALMA: continuum at 1.3 mm and molecular lines retrieved from the ALMA data archive (Project 2015.1.01312.S; PI: Fuller, G.; Band 6). The angular and spectral resolution are 0$.\!\!^{\prime\prime}$78$\times$0$.\!\!^{\prime\prime}$60~and 1.4 km s$^{-1}$, respectively. The radio continuum observations and the millimeter data were handled with CASA version 5.8.0-105.

Finally, we carried out IFU spectroscopic NIR observations using NIFS at Gemini North (project GN-2022A-Q-125; PI: S. Paron). The central wavelength of these observations is $\lambda_{c} = 2.20~\mu$m, while the spectral range and resolution are 1.99-2.40 $\mu$m and 5290, respectively\footnote{https://www.gemini.edu/instrumentation/nifs}. Four fields that uncover the whole region of G29 were observed. The standard NIFS tasks included in the Gemini IRAF package v1.14 were used for data reduction.

In what follows, we present some preliminary results obtained from the mentioned multifrequency dataset.

\section{Results}\label{res}

\subsection{New molecular species in YSO-G29}

Upon examining the data cubes from the ALMA observations, we identified numerous molecular emission lines, revealing the presence of many chemical species in the region.

\begin{figure*}[h!]
\centering
\includegraphics[width=15cm]{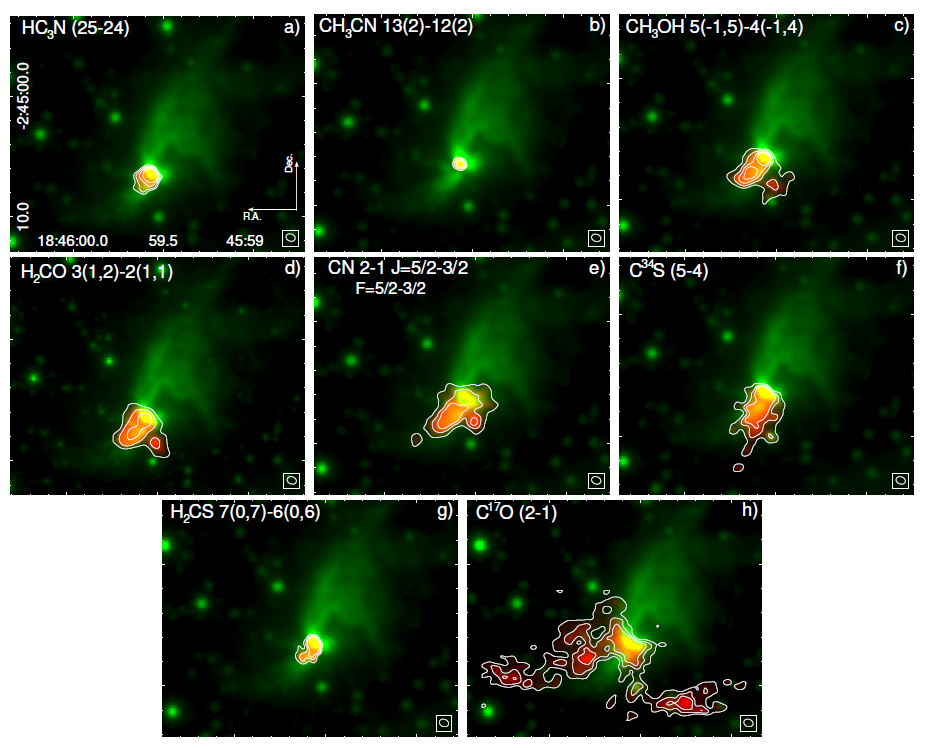}   
\caption{Integrated emission maps of the molecules listed in Table\,\ref{table} (molecular emission in red with white contours) superimposed on the Ks-emission obtained with Gemini-NIRI (green). The molecular lines were integrated between 95 and 105 km s$^{-1}$. In all cases, the first contour represents emission at level of 3$\sigma$. The beam of the molecular observations is included at the bottom right corner of each panel.}
\label{maps}
\end{figure*}

Table\,\ref{table} summarizes only eight of such molecules. Detecting these new molecular species contributes to overview the chemical richness in the region. Figure\,\ref{maps} displays maps of the molecular line emission integrated between 95 and 105 km~s$^{-1}$ (presented in red with white contours), superimposed to the Ks-emission (in green).

All new chemical species found in this work, and their spatial distribution can help us to study the star-forming processes occurring in this region. We discuss those listed in Table\,\ref{table}. The C$^{34}$S maximum emission (see Fig.\,\ref{maps}f) coincides with the peak of continuum emission at 1.3 mm and extends far to the south. The CN species (Fig.\,\ref{maps}e) exhibits a similar morphology but is shifted toward the southeast relative to the C$^{34}$S emission, revealing more diffuse regions around such core. Both species are tracers of cavities generated by outflows \citep{ortega23}, and in this case, by comparing their spatial distribution, it is worth noting that they present slightly different orientations.
In Fig.\,\ref{maps}h, the emission of C$^{17}$O exhibits a widespread distribution throughout the entire analyzed region, probably featuring the areas of the outer envelope where the core is embedded. The HC$_{3}$N, CH$_{3}$CN, and H$_{2}$CS (Fig.\,\ref{maps}a, b, and g, respectively) present a compact emission, outlining the innermost region of the core. Both CH$_{3}$OH and H$_{2}$CO have emissions concentrated in the core detected in the continuum at 1.3 mm and extend southeastward, probably also tracing outflow activity.

\begin{figure}[h!]
    \centering
    \includegraphics[width=7.5cm]{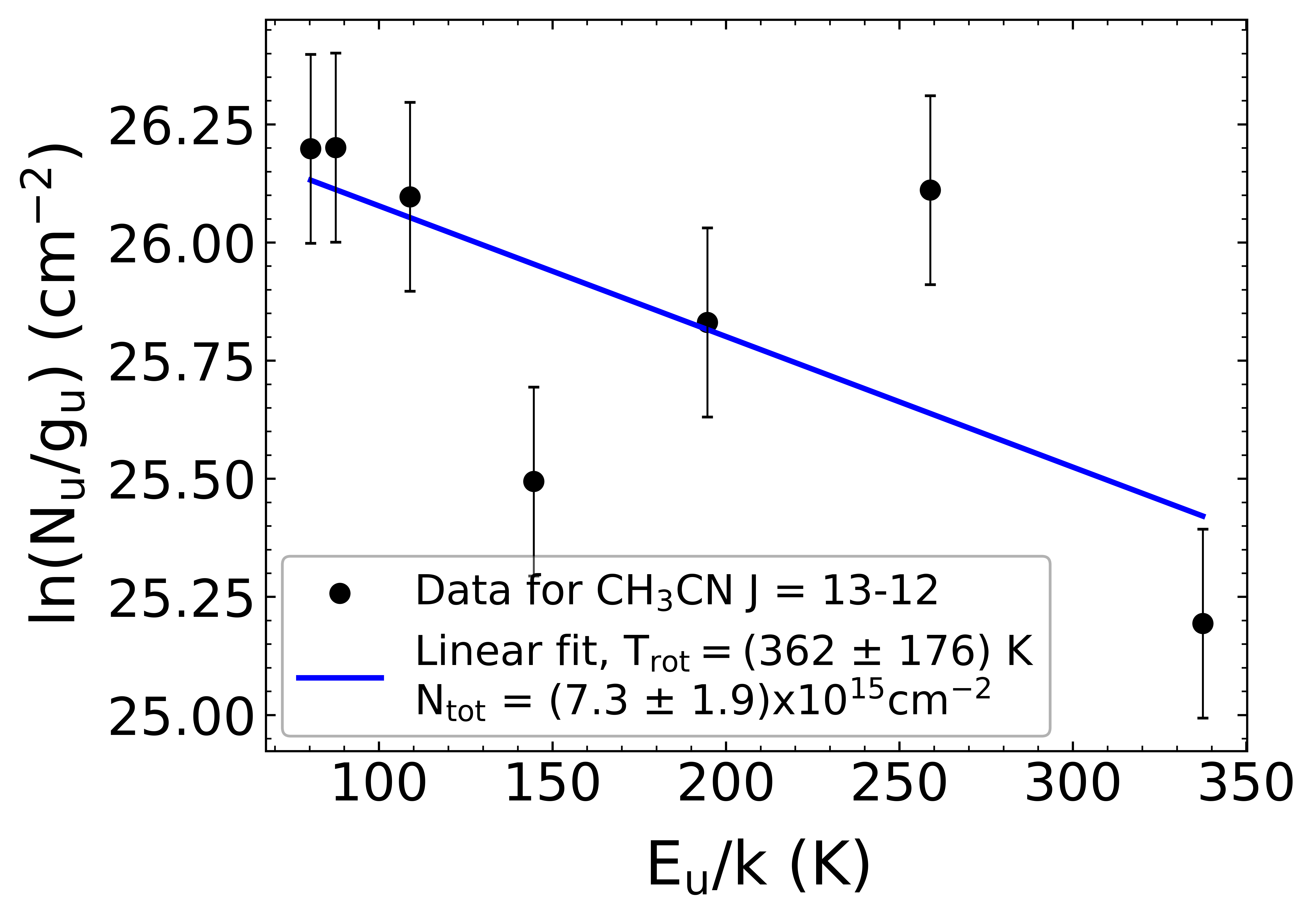}
    \caption{Rotational diagram for CH$_{3}$CN \textit{J} = 13--12 transition, based on the \textit{K} = 0 to 6 projections. The blue line represents the best linear fit of the data.}
    \label{trot}
\end{figure}

\begin{table}
\caption{Analyzed molecular lines.}
\centering
\tiny
\begin{tabular}{lcc}
\hline\hline\noalign{\smallskip}
Molecule    & Transition &  Rest Freq.(GHz)$^{1}$        \\
\hline\noalign{\smallskip}
C$^{17}$O   & 2--1            & 224.714               \\             
H$_{2}$CO   & 3(1,2)--2(1,1)            & 225.697               \\             
CN   & 2--1 J=5/2--3/2 F=5/2--3/2       & 226.874               \\             
HC$_{3}$N   & 25--24            & 227.418               \\             
CH$_{3}$CN           & 13--12 k=2 & 239.119           \\
H$_{2}$CS           & 7(0,7)--6(0,6) & 240.267           \\
C$^{34}$S           & 5--4 & 241.016           \\
CH$_{3}$OH                & 5(-1,5)--4(-1,4)         & 241.767                    \\
\hline

\multicolumn{3}{l}{$^{1}$Rest frequencies from the \href{https://splatalogue.online}{Splatalogue Catalog}.}\\
\end{tabular}
\label{table}
\end{table}

We estimated the rotational temperature of the core (T$\rm_{rot}$) by using the population diagram method \citep{goldsmith99} applied to the CH$_{3}$CN \textit{J} =13--12 (\textit{K} = 0 to 6 projections). Assuming optical thin lines and Local Thermodynamic Equilibrium (LTE) conditions, we employed the following equation to estimate T$\rm_{rot}$:

\begin{align}
{\rm ln(N_u/g_u}) = {\rm ln(N_{tot}/Q_{rot})-(E_u/kT_{rot})}
     \label{RotDia}    
\end{align}

\noindent
where ${\rm N_u}$ represents the molecular column density of the upper level of the transition, ${\rm g_u}$  the total degeneracy of the upper level, ${\rm E_u}$ the energy of the upper level, ${\rm N_{tot}}$ the total column density of the molecule, ${\rm Q_{rot}}$ the rotational partition function, and k the Boltzmann constant. For interferometric observations, the left-hand side of Eq.\,\ref{RotDia} can be rewritten with parameters associated with the beam size, the frequency transition, and integrated intensity (see details in \citealt{ortega23}, Sect.\,4.3.1).
We performed a linear fitting to Eq.\,\ref{RotDia} (see Fig.\ref{trot}), and from the slope, we determined ${\rm T_{rot}}$. N$\rm_{tot}$ was obtained from the other term of the equation using the Q$\rm_{rot}$(T$\rm_{rot}$) value from the CDMS database\footnote{https://cdms.astro.uni-koeln.de/cdms/portal/queryForm}. This procedure yields T$_{\rm rot} = 362\pm176$ K and N$_{\rm tot}$ $ = (7.3\pm1.9)\times10^{15}$ cm$^{-2}$.  

\subsection{Ionized gas in the region}

\begin{figure}
    \centering
    \includegraphics[width=8.65cm]{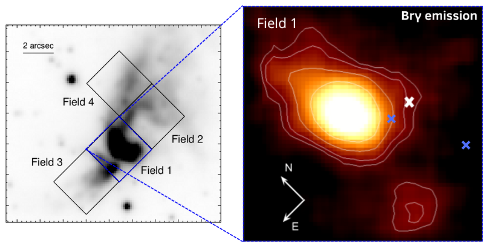}
    \caption{\textit{Left:} Ks-band emission toward G29. The black boxes are the observed fields using Gemini-NIFS. \textit{Right:} Preliminary image obtained from the Gemini-NIFS data towards Field\,1 in which the integrated Br$\gamma$ emission is displayed. The image is not flux-calibrated and the contours are just to remark the emission gradient. The white cross indicates the position of the radio continuum source observed at 3 cm with JVLA, and the blue crosses represent the peaks of the continuum at 1.3 mm obtained with ALMA (black contours in Fig.\,\ref{g29}).}
    \label{gemini}
\end{figure}

Figure\,\ref{gemini} (right) displays a preliminary image of the integrated Br$\gamma$ line (at 2.1686 $\mu$m) observed with Gemini-NIFS toward the center of G29 (Field\,1; see Fig.\ref{gemini} left). As a reference, the position of a compact radio continuum source at 3 cm, which we discovered using the JVLA observations described above, is indicated with the white X. The analysis of this source will be presented in a future work (Martinez et al., in preparation). Here, we simply show the position of this bright, compact radio source, which has a size slightly larger than the beam.
The blue crosses indicate the ALMA continuum peaks at 1.3 mm from Fig.\,\ref{g29}. Other spectral line maps (obtained with Gemini-NIFS) in the NIR (H$_{2}$ emission lines and continuum emission) are being processed.

\section{Discussion}

The analysis of the molecular emission, as presented here, and the involved chemistry, is very useful for understanding the physical processes occurring in star formation sites and the conditions there present. In this context, our focus is on elucidating the chemistry within G29 region.

The rotational diagram method sheds light on certain physical parameters of the region, for instance, molecular column density and temperature. The CH$_{3}$CN column density was derived to be $(7.3\pm1.9)\times10^{15}$ cm$^{-2}$ and the rotational temperature of the core was calculated as $362\pm176$ K in agreement with the typical temperatures observed in hot cores (e.g. \citealt{ortega23}). The temperature of the region was previously assessed by \cite{areal20} employing the same methodology, albeit with only four CH$_{3}$CN {\it K} projections. In that specific instance, T$\rm_{rot}$ was significantly lower than the current result regarding the temperature expected for a hot molecular core. 
The updated value presented herein extends upon the findings reported in the previous work, and it is deemed more precise in characterizing the temperature of the region.

Both $\rm CH_{3}CN$ and $\rm CH_{3}CCH$ are commonly found in hot cores; hence, their presence is anticipated within the core region since temperatures greater than 150 K allow the evaporation of the icy molecular mantles in the dust grains and enrich the gas around the protostar. On the other hand, the identified $\rm H_{2}CO$ transition has an upper-level energy of about 33 K 
signifying its capability to trace the colder parts of the molecular core envelopes \citep{dima}.

Since the production of H$_{2}$CS arises from the transformation of SO and SO$_{2}$ in grain mantles, the presence of this molecule accounts for a colder envelope, from where it is released into the gas phase by successive heating around a YSO \citep{minh}. Nevertheless, in this case, we suggest that the occurrence of C$^{34}$S in the region is not merely related to the destruction of H$_{2}$CS (through electronic recombination of H$_{3}$CS$^{+}$, as described by \citealt{charnley}). Assuming that the C$^{34}$S
species is present in a solid state on the surface of dust
grains, its enhancement in the gas phase
might be attributed to a shock effect, likely induced by
outflow activity. This would explain the distinctive spatial distribution depicted in Fig.\,\ref{maps}.

It is established that CH$_{3}$OH undergoes sublimation from icy dust mantles when encountering turbulent gas induced by outflow passages \citep{disho}. Additionally, CN species could trace the cavities generated by the outflows \citep{ortega23}. We suggest that both species reveal the direction of an outflow that is not coincident in direction with the one described by the C$^{34}$S emission.

The preliminary analysis of the JVLA data indicates the presence of a new source in the G29 region which could contribute to explain, for instance, the complex morphology of the NIR emission. In an upcoming study, we will analyze how this discovery affects the chemistry in the region and whether it is responsible for the different features observed. Among them are ionization processes, which can significantly contribute to ion-neutral chemistry and can photodissociate molecular species. In this context, the presence of the compact radio source and the large Br$\gamma$ feature (see Fig.\,\ref{gemini}) can provide insightful information about the chemistry at the central region. 


\begin{acknowledgement}
N.C.M. is a doctoral fellow of CONICET, Argentina. This work was partially supported by the Argentinian grants PIP 2021 11220200100012
and PICT 2021-GRF-TII-00061
awarded by CONICET
and ANPCYT.
\end{acknowledgement}


\bibliographystyle{baaa}
\small
\bibliography{bibliografia}

\end{document}